\documentclass{elsart}

\usepackage{graphicx}

\usepackage{amssymb}

\begin{document}

\begin{frontmatter}



\title{Theoretical Aspects of Neutrino Oscillations\thanksref{label1}}
\thanks[label1]{Invited Talk at the 
NuFACT'01 Conference, Tsukuba, Japan, 24--30 May 2001.}


\author{Andr\'e de Gouv\^ea\thanksref{label2}}
\thanks[label2]{New address from September 2001: Theoretical Physics 
Devision, Fermilab, MS 106, P.O. Box 500, Batavia, IL 60510, USA.}
\address{CERN - TH Division, CH-1211, Gen\`eve 23, Switzerland}

\begin{abstract}
I review some aspects concerning the physics of neutrino mixing and oscillations. 
I discuss in some detail the physical neutrino oscillations parameter 
space in the case of two and three family mixing, and briefly describe the current 
knowledge of neutrino mixing parameters according to the present solar, atmospheric, 
and reactor neutrino data. I also briefly comment on the possibility of solving
the LNSD anomaly together with the solar and atmospheric ones. 
I conclude by emphasising that that even though in five 
to ten years time a lot will be learnt from the next round of neutrino experiments, 
a great deal about neutrino masses and neutrino mixing will remain unknown. 
\end{abstract}

\end{frontmatter}

\section{Introduction and Motivation}
\label{intro}
There are, currently, three neutrino puzzles, two which strongly indicate the presence
of physics beyond the standard model (SM).
The oldest one -- the solar neutrino puzzle \cite{Solar_talk}-- 
is the fact that the flux of electron-type solar
neutrinos measured at the Earth is significantly smaller than predicted by 
solar physics models. This deficit is confirmed by 
different experiments, which make use of very different techniques for 
detecting neutrinos. It is fair to say that, after the publication of the first 
results from SNO \cite{SNO}, there is very strong evidence for a flux of $\nu_{\mu,\tau}$ 
coming from the Sun.  

The second -- the atmospheric neutrino puzzle \cite{Atm_talk} -- is the fact that the flux
of atmospheric muon-type neutrinos and antineutrinos differs significantly
from theoretical predictions. 
This anomalous behaviour was first observed by the 
proton decay experiments IMB and Kamiokande, and later confirmed and firmly 
established by the SuperKamiokande experiment. The anomaly manifests itself more
strongly in the ratio of the predicted $\nu_e$ and $\nu_{\mu}$ flux-ratio to the
experimentally measured one, and the nontrivial angular dependency of the
$\nu_{\mu}$ flux. The latter is by far the most striking evidence for physics beyond
the SM we have at the moment.     

Finally, the LSND experiment \cite{LSND}, 
which studies neutrinos produced after pion and muon 
decays, has observes a 3-sigma excess of $\bar{\nu}_e$-like events from $\mu^{+}$
decays. The LSND anomaly has not been confirmed or excluded by KARMEN \cite{KARMEN}
or other neutrino experiments. 
The situation will improve significantly with the advent of the MiniBoone
\cite{MiniBoone} experiment, which is due to start taking data in 2002. 

The neutrino puzzles are best solved by assuming that the neutrinos have mass,
and that neutrino mass eigenstates and weak eigenstates differ, hence the neutrinos 
``oscillate.'' The solar neutrino puzzle is best solved by assuming 
$\nu_e\leftrightarrow\nu_{\mu,\tau}$ oscillations \cite{Carlos}, 
while the atmospheric neutrino
puzzle hints at quasi-maximal $\nu_{\mu}\leftrightarrow\nu_{\tau}$ oscillations 
\cite{Lisi}. The LSND anomaly requires $\bar{\nu}_{\mu}\leftrightarrow\bar{\nu}_e$ 
oscillations. There are,
however, other exotic solutions to individual neutrino puzzles in the market, such as
new neutrino--matter interactions, violation of Lorentz invariance and quantum 
mechanics, and neutrino decays, to name a few. Usually, these solutions are tailor-made 
to address a particular neutrino puzzle. It is fair to say that neutrino oscillations
is the only hypothesis that can properly address {\sl all} neutrino anomalies. 

In this talk, I review some aspects of neutrino oscillations. In the next section,
I discuss neutrino mixing and how it leads to neutrino flavour conversion, both in the
absence and presence of a medium. I concentrate on two and three family mixing
scenarios, and spend some time discussing the ``physical parameter space.'' 
In Sec.~\ref{sec:at_work}, I briefly review how neutrino oscillations solve the neutrino 
puzzles, and what the current data can and 
cannot tell us about the oscillation
parameters. In Sec.~\ref{sec:LSND}, I discuss some of the issues that must be faced if 
the LSND anomaly is also to be explained by neutrino oscillations, and in 
Sec.~\ref{sec:outlook} I 
conclude with an outlook regarding what we may expect to learn in the next few years and 
what will be left to do with future, more powerful, machines (neutrino factories?).  

\section{Neutrino Mixing and Oscillations}
\label{sec:mixing}

If neutrinos have mass, there is no reason for the mass eigenstates to coincide with
the weak (also referred to as flavour) eigenstates. This not being the case, the
two bases are connected by a unitary matrix
$\nu_{\alpha} = U_{\alpha i}\nu_i$,
where $\nu_{\alpha}$ are weak eigenstates ($\alpha=e,\mu,\tau,\dots$) and $\nu_i$ are
mass eigenstates, with masses $m_i$ ($i=1,2,3,\dots$). 
$U_{\alpha i}$ will be
referred to as the neutrino mixing matrix.\footnote{There is some controversy regarding
how this matrix should be called. I will stir clear of this issue here.} 
Note that one can choose
to label the mass eigenstates {\sl in ascending order of mass-squared,}\/ with no loss
of generality. This will be assumed, unless otherwise noted.

When neutrinos propagate in vacuum, it is very simple to calculate the probability
that a neutrino with energy $E_{\nu}$, 
which is produced as a weak eigenstate $\nu_{\alpha}$, is detected
a distance $L$ from the source as a weak eigenstate $\nu_{\beta}$:
\begin{eqnarray}
P(\nu_{\alpha}\rightarrow \nu_{\beta})(L)\equiv P_{\alpha\beta} & = &  
\delta_{\alpha\beta} - 4\sum_{i<j} \Re(U_{\alpha i}U_{\alpha j}^*U_{\beta i}^*U_{\beta j})
\sin^2\left(\frac{\Delta m^2_{ij} L}{4E_{\nu}}\right) \nonumber \\ & - &
2\sum_{i<j} \Im(U_{\alpha i}U_{\alpha j}^*U_{\beta i}^*U_{\beta j})
\sin\left(\frac{\Delta m^2_{ij} L}{2E_{\nu}}\right),
\label{Pab}
\end{eqnarray}
where $\Delta m^2_{ij}\equiv m_j^2-m_i^2$ are the mass-squared differences. For 
antineutrinos, one simply has to replace $U_{\alpha i}\leftrightarrow U_{\alpha i}^*$,    
meaning that eq.~(\ref{Pab}) holds as long as the sign of the last term is flipped.
One can define the different oscillation lengths
\begin{equation}
L_{\rm osc}^{ij} = \pi \frac{4E_{\nu}}{\Delta m^2_{ij}} = \pi \left(\frac{E_{\nu}}{\rm GeV}
\right)\left( 
\frac{\rm eV^2}{1.267 \Delta m^2_{ij}}\right) \rm [km]. 
\end{equation}

\begin{figure}
\centerline{
\parbox{0.62\textwidth}{\includegraphics[width=0.62\textwidth]{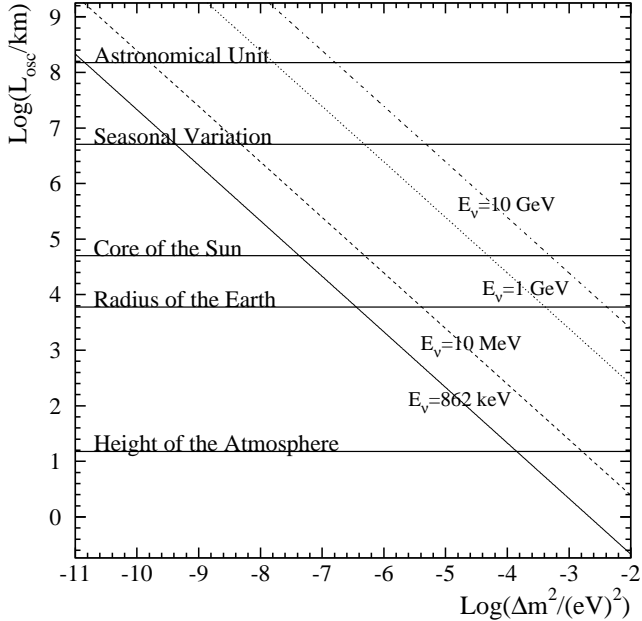}}
\parbox{0.38\textwidth}{\caption{$L_{\rm osc}^{ij}$ as a function of $\Delta m^2_{ij}$,
for different values of the neutrino energy, namely, 
$E_{\nu}$=862~keV ($^7$Be solar neutrino), $E_{\nu}$=10~MeV (``typical''
$^8$B solar neutrino), $E_{\nu}$=1~GeV (``typical'' atmospheric neutrino) and 
$E_{\nu}$=10~GeV.}}
\label{explain}}
\end{figure}

Figure~\ref{explain} depicts $L_{\rm osc}^{ij}$ as a function of $\Delta m^2_{ij}$
for different values of $E_{\nu}$.
When the neutrinos propagate in matter (of either constant or varying density) 
the situation is significantly altered, as will be discussed in the next two subsections.

\subsection{Two Neutrino Mixing}

If there are oscillation between only two neutrino states, 
the situation is rather simple. The neutrino mixing
matrix is parametrised in terms of only one angle $\theta$, and one extra phase if
the neutrinos are Majorana particles. This phase, however, is not observable
in oscillation phenomena, and may be safely set to zero. Explicitly, if there is
mixing only between $\nu_e$ and $\nu_x$,
\begin{equation}
\left(\matrix{\nu_{e} \cr \nu_{x} }\right)=
\left(\matrix{\cos\theta&\sin\theta\cr
-\sin\theta&\cos\theta}\right) \left(\matrix{\nu_{1} \cr \nu_{2} }\right),
\end{equation}
and
\begin{equation}
P_{ee}^{\rm vac}=1-\sin^22\theta
\sin^2\left(\frac{\Delta m^2_{12}L}{4E_{\nu}}\right).
\label{P_2nus}
\end{equation}

In the presence of neutrino--medium interactions, the neutrino propagation is modified
\cite{review}.
These effects can be expressed in terms of a ``matter potential,'' such that 
neutrino propagation is governed by the following Sch\"odinger-like equation
(in the weak basis),
\begin{equation}
i\frac{\rm d}{{\rm d}L}
\left(\matrix{\nu_{e} \cr \nu_{x} }\right)=\left[\frac{\Delta m^2_{12}}{2E_{\nu}}
\left(\matrix{\sin^2\theta&\cos\theta\sin\theta\cr
\cos\theta\sin\theta&\cos^2\theta}\right)+
\left(\matrix{V_{ee}&V_{ex}\cr
V_{ex}^*&V_{xx}}\right)
\right] \left(\matrix{\nu_{e} \cr \nu_{x} }\right),
\label{de_2}
\end{equation}
where $V_{\alpha\beta}$ are the different matter potential functions, which are in principle
a function of the position. If $\nu_{x}$ is a linear combination of the other active
neutrinos ($\nu_{\mu}$ and $\nu_{\tau}$), the matter potential matrix is such that
$V_{ex}=0$ and $V_{ee}-V_{xx}=\sqrt{2}G_{F}N_e$,\footnote{Note that the ``Hamiltonian''
in eq.~(\ref{de_2}) is defined up to a term proportional to the identity matrix, which 
does not yield any physically observable effect.} 
if the neutrinos propagate in ``normal''
matter with electron number density $N_e$, 
and only standard model (SM) interactions are taken into account. For antineutrinos, $N_e$
is replaced by $-N_{e}$ (which is the positron number density). 

Two important issues can be addressed just by inspecting the general form of 
eq.~(\ref{de_2}). First of all, while eq.~(\ref{de_2})
is, in principle, not solvable, we can still 
construct useful relations between the different $P_{\alpha\beta}$. In particular,
because the ``Hamiltonian'' is hermitian (this condition would break down, for
example, in the presence of neutrino absorption by the medium or if the neutrinos decay),
$\sum_{\alpha}P_{\alpha\beta}=\sum_{\beta}P_{\alpha\beta}=1$. This implies, in the 
case of two neutrino oscillations, that only one $P_{\alpha\beta}$ is independent. 
For example, one can express all $P_{\alpha\beta}$ in terms of $P_{ee}$: 
$P_{ex}=P_{xe}=1-P_{ee}$, $P_{xx}=P_{ee}$. It is curious to note that, in the case
of two family oscillation, $P_{ex}=P_{xe}$ (T-invariance) is guaranteed, while
nothing, in principle, prevents $P_{ee}\neq P_{\bar{e}\bar{e}}$ (matter induced
CP-violation). 

Second, it is interesting to define what is the ``physical range'' for the oscillation
parameter $\theta$, {\it i.e.}\/ what are the values of $\theta$ which yield, if one
measures $P_{\alpha\beta}(L)$, distinct results? Initially, because $\theta$ is an
angle, it need be defined only from 0 to $2\pi$. On top of this, it is trivial to 
check that the transformation $\theta\rightarrow \pi-\theta$ is equivalent to
$\theta\rightarrow -\theta$, in the sense that the differential equations eq.~(\ref{de_2})
transform in exactly the same way. This implies that choosing values of $\theta=[0,\pi]$ is 
enough (this may also be seem by noting that eq.~(\ref{de_2}) depends only on 
$2\theta$, up to irrelevant terms proportional to the unit matrix). 
Furthermore, {\sl if the matter potential matrix is diagonal,}\/ the 
transformation $\theta\rightarrow \pi-\theta$ modifies the differential equations in such a
way that $(\nu_{e}(L),\nu_{x}(L))_{\pi-\theta}=(\nu_{e}(L),-\nu_{x}(L))_{\theta}$. Because
$P_{\alpha\beta}(L)=|\nu_{\beta}(L)|^2$ assuming the boundary condition is a pure
$\nu_{\alpha}$ state at $L=0$, $\theta$ and $\pi-\theta$ yield the same 
$P_{\alpha\beta}(L)$, and the physical parameter space can be safely restricted to 
$[0,\pi/2]$. This is {\sl not} that case if the matter potential has off-diagonal terms
and/or if the ``weak'' eigenstates at the production point differ from the ``weak''
eigenstates at the detection point. This is not the case within the SM, 
but in extensions to it one may indeed be able to distinguish $\theta$ from 
$\pi-\theta$.
Finally, in the absence of a matter potential (which is the case of vacuum oscillations), 
there is yet another ``symmetry:'' $(\nu_{e}(L),\nu_{x}(L))_{\pi/2-\theta}=
(\nu_{e}^*(L),-\nu_{x}^*(L))_{\theta}$, such that the entire physical parameter space
can be spanned by allowing $\theta$ to vary from 0 to $\pi/4$. This can be explicitly
seen in eq.~(\ref{P_2nus}).       

The parameter space in the absence of physics beyond the SM is easy to interpret. In
the region $\theta=[0,\pi/4]$ (the light side) $\nu_e$ is ``predominantly 
light'' ($\nu_1$), while when $\theta=[\pi/4,\pi/2]$
(the dark side) $\nu_e$ is ``predominantly heavy'' ($\nu_2$). While pure vacuum 
oscillations cannot tell the dark from the light side, matter effects brek the degeneracy
and allow one to explore the entire $[0,\pi/2]$ range.

In many cases of interest, eq.~(\ref{de_2}) can be solved, at least approximately. If the
neutrinos propagate in a constant electron number density, $P_{ee}$ is easily 
calculable, and is given by eq.~(\ref{P_2nus}), where the mass-squared difference and
the mixing angle are modified to ``matter'' parameters, given by
\begin{eqnarray}
\label{costhm}
\cos2\theta_{M}&=&\frac{\Delta_{12}\cos2\theta-A}{\sqrt{\Delta_{12}^2+A^2
-2A\Delta_{12}\cos2\theta + 4\Re(V_{ex})\Delta_{12}\sin2\theta+4|V_{ex}|^2}}, \\
\Delta_{12}^M &=& \sqrt{\Delta_{12}^2+A^2-2A\Delta_{12}\cos2\theta
+4\Re(V_{ex})\Delta_{12}\sin2\theta+4|V_{ex}|^2},
\end{eqnarray}
where $\Delta_{ij}\equiv \Delta m^2_{ij}/(2E_{\nu})$ and $A\equiv{V_{ee}-V_{xx}}$. 
As discussed in the previous paragraph, the matter angle and frequency  
depend only on $2\theta$, and, if $V_{ex}=0$, are invariant under 
$\theta\rightarrow\pi-\theta$.

Another very useful solution exists in the case of neutrinos propagating in a monotonically
falling electron number density, as is the case of neutrinos produced in the Sun's 
core \cite{Petcov_1}:
\begin{equation}
P_{ee}=P_1 \cos^2\theta + P_2 \sin^2\theta -\cos 2\theta_M 
\sqrt{P_c(1-P_c)}\sin2\theta\cos(\Delta_{12}L+\phi_M), 
\label{P_solar}
\end{equation}
where $P_{1}=1-P_2=1/2+1/2(1-2P_c)\cos 2\theta_M$, $P_c$ is the 
``level crossing probability,'' $\phi_M$ is a constant matter-induced phase,
and the matter angle is to be computed at the production point. If the neutrino
propagation inside the Sun is adiabatic, $P_c=0$, while in the case of an exponential
electron number density profile $N_e(x)=N_0\exp(-x/r_0)$, the approximate form
$P_c=(\exp(-\gamma\sin^2\theta)-\exp(-\gamma))/(1-\exp(-\gamma))$, where
$\gamma=2\pi r_0\Delta_{12}$, is valid in a large portion
of the parameter space \cite{Petcov_2}. 
Recently, a lot of progress has been made concerning the 
understanding of solar neutrino oscillations, including studies of whether the
expressions above are valid for $\theta>\pi/2$ \cite{day_night,alex1} 
(the dark side of the parameter space), 
and how they should be understood and interpreted \cite{alex2}. 

\subsection{Three Neutrino Mixing}

The three by three neutrino mixing matrix which connects $\nu_{e}$, $\nu_{\mu}$, and
$\nu_{\tau}$ to the mass eigenstates $\nu_{1,2,3}$ is parametrised by three angles and
one complex phase, plus two Majorana phases which are not detectable via neutrino 
oscillation experiments and can be safely set to zero. It is ``traditional'' to define
the mixing angles $\theta_{12,13,23}$ in the following way:
\begin{equation}
\tan^2\theta_{12}\equiv \frac{|U_{e2}|^2}{|U_{e1}|^2},~~~
\tan^2\theta_{23}\equiv \frac{|U_{\mu3}|^2}{|U_{\tau3}|^2},~~~
\sin^2\theta_{13}\equiv |U_{e3}|^2,
\end{equation} 
while 
\begin{equation}
\Im(U_{e2}^*U_{e3}U_{\mu2}U_{\mu3}^*)\equiv 
\sin\theta_{12}\cos\theta_{12}\sin\theta_{23}\cos\theta_{23} 
\sin\theta_{13}\cos^2\theta_{13}\sin\delta 
\end{equation}
defines the CP-odd phase $\delta$. 

In the case of pure vacuum oscillations, the survival probabilities are given
by eq.~(\ref{Pab}). It should be readily noted that, unlike the
two-family case, if $\delta\neq 0$ or $\pi$, 
$P_{\alpha\beta}\neq P_{\bar{\alpha}\bar{\beta}}$, signalling CP-violation in the
neutrino sector.\footnote{Of course, there is also T-violation,
 $P_{\alpha\beta}\neq P_{\beta\alpha}$, such that 
$P_{\alpha\beta}= P_{\bar{\beta}\bar{\alpha}}$ (CPT).}

If the neutrinos propagate inside a medium, the situation is (much) more complicated.
In principle, one has to solve the Schr\"odinger-like equation (in the weak basis)
\begin{equation}
i\frac{\rm d}{{\rm d}L}\nu_{\alpha}=\left[\sum_{i=2,3}
\left(\frac{\Delta m^2_{1i}}{2E_{\nu}}\right)
U^*_{\alpha i}U_{\beta i}
+ V_{\alpha\beta}\right]\nu_{\beta}, ~~{\rm where}~ V_{\alpha\beta}=V_{\beta\alpha}^*.
\label{de_3}
\end{equation}
On the other hand,
because the ``Hamiltonian'' in eq.~(\ref{de_3}) is hermitian, the different 
oscillation probabilities are related. As before, the constraints 
$\sum_{\alpha}P_{\alpha\beta}=1$ and $\sum_{\beta}P_{\alpha\beta}=1$ arise from the
unitary evolution of the quantum state. In the case of three family oscillation, they
imply that there are only four independent $P_{\alpha\beta}$, which one may choose to
be, say, the three ``diagonal'' $P_{\alpha\alpha}$ plus $P_{e\mu}$. 
The other five are linear 
combinations of these four. Explicitly,
\begin{eqnarray}
&P_{e\tau}=1-P_{ee}-P_{e\mu},~~~~~~~P_{\tau e}=P_{\mu\mu}+P_{e\mu}-P_{\tau\tau}, \nonumber \\
&P_{\mu e}=1+P_{\tau\tau}-P_{ee}-P_{\mu\mu}-P_{e\mu}, \\
&P_{\mu\tau}=P_{ee}+P_{e\mu}-P_{\tau\tau},~~~~~~~P_{\tau\mu}=1-P_{\mu\mu}-P_{e\mu}. \nonumber
\end{eqnarray}
Note that, unlike the two-family case, 
$P_{\alpha\beta}$ may differ from $P_{\beta\alpha}$, even if the CP-odd phase $\delta$ is 
``turned off'' \cite{GeV_solar}. 
This means that the presence of the medium can induce not only
matter CP-violation, but all matter T-violation. In order for this to happen, all that
is required is that some $V_{\alpha\beta}$ breaks T-invariance. This does not happen
in the presence of a constant matter density, but may certainly happen, for example,
for neutrinos produced inside the Sun \cite{GeV_solar}. Whether or not such effects are
relevant for terrestrial neutrino beams has also been studied \cite{AL}.

Also from eq.~(\ref{de_3}), it is possible to determine what is the ``physical range''
for the parameters $\theta_{ij}$ and $\delta$. One can directly check, as discussed in the
two-family case, that if $V_{\alpha\beta}=0$ for $\alpha\neq\beta$, 
$V_{\mu\mu}=V_{\tau\tau}$, and the weak 
eigenstates are the same at the production and detection locations
(as are the conditions in ``normal'' matter, including no interactions beyond the SM), 
all physically distinguishable values for the oscillation probabilities are 
probed if the three
angles are allowed to vary in the range $[0,\pi/2]$, while $\delta$ covers the full range 
$[-\pi,\pi]$. Note that, unlike the two-flavour case,  
there is no further reduction of the parameter space if the matter potential vanishes.    
In the presence of new physics effects that lead to a non-diagonal matter potential,
a larger parameter space is, in principle,
required in order to describe all possible oscillation scenarios. 

One final convention issue should be mentioned. Thanks to the current experimental data,
it has become customary to redefine the order of the mass eigenstates if 
$\Delta m^2_{12}>\Delta m^2_{23}$. Whenever this (an ``inverted hierarchy'') 
happens, the mass-eigenstates are
relabelled $1 \rightarrow 3 \rightarrow 2$, such that $\Delta m^2_{13,23}<0$ and
$\Delta m^2_{12}<|\Delta m^2_{23}|$, always. This redefinition is done such that one
can relate $\theta_{12}$ with the ``solar'' angle, $\theta_{23}$ with the 
``atmospheric'' angle, and $\theta_{13}$ with the ``reactor'' angle, as will be
discussed briefly in the next section. Note that within this 
definition of the mass eigenstates ({\it i.e.,}\/ $m_1^2<m^2_2$ and $\nu_3$ is defined 
such that $\Delta m^2_{12}<|\Delta m^2_{23}|$), 
the sign of $\Delta m^2_{23}$ becomes ``physical,'' and determines whether the neutrino 
masses are hierarchical, or whether $m_1^2,m^2_2$ are quasi-degenerate 
($\Delta m^2_{12}\ll m^2_2$) in an inverted hierarchy. It should be
noted that oscillation experiments
cannot distinguish whether all three mass eigenstates are quasi-degenerate or not. 

Since we are now dealing with three-by-three matrices, simple exact solutions to 
eq.~(\ref{de_3}) do not exist, even in the case of a constant matter potential 
(see, however, \cite{Freund}). However,
Nature seems to have been kind, and a few approximations seem to be well justified,
including $\Delta m^2_{12}/|\Delta m^2_{23}|\ll 1$\footnote{see talk by Eligio 
Lisi \cite{Lisi} concerning the violation of this condition.} 
and $\sin^2\theta_{13} \ll 1$. Under
these circumstances many of the well known two neutrino results can be applied. A good
rule of thumb is that, in the presence of ``normal'' matter (no new physics), 
$\theta_{13}$ is modified to a 
matter $\theta_{13}^M$, given by eq.~(\ref{costhm}) with $\theta\rightarrow \theta_{13}$
and $\Delta_{12}\rightarrow\Delta_{13}$; $\theta_{12}$ is also modified to a matter angle
as defined in eq.~(\ref{costhm}) with $\theta\rightarrow \theta_{12}$ and a modified
matter potential: $A\rightarrow A\times \cos^2\theta_{13}$; $\theta_{23}$ and
$\delta$ are not modified. For solar electron-type 
neutrinos (with $E_{\nu}\lesssim 10$~MeV), for example, 
\begin{equation}
P_{ee}^{3\nu}=\cos^4\theta_{13}P_{ee}^{2\nu}(\theta\rightarrow\theta_{12}, 
A\rightarrow A\cos^2\theta_{13}) + \sin^4\theta_{13},
\label{Psol_3}
\end{equation}
where $P_{ee}^{2\nu}$ is given
by eq.~(\ref{P_solar}), and $\Delta_{13}$ effects are ``averaged out'' 
(see \cite{review} and references therein).  

\section{Neutrino Oscillations in Action}
\label{sec:at_work}

Here, I briefly review what the experimental data has to say about the neutrino 
oscillation parameters if the neutrino puzzles are interpreted as evidence
for neutrino oscillations. As mentioned in the introduction, 
I will ignore the yet unconfirmed LSND anomaly, and
concentrate on addressing only the solar and atmospheric neutrino data, while satisfying
the constraints imposed by reactor neutrino experiments. 

First of all, reactor neutrino experiments \cite{reactor_exp}
have, so far, failed to observe a depletion
of the $\bar{\nu}_e$-flux produced by nuclear reactors. The survival probability
\begin{equation}
P_{\bar{e}\bar{e}} = 1-\sin^22\theta_{13}\sin^2\left(\frac{\Delta_{23}L}{2}\right) 
+ O(\Delta_{12}L),
\end{equation}
and, 
in light of the atmospheric and solar data, $P_{\bar{e}\bar{e}} \sim 1$ implies
$\sin^22\theta_{13}\lesssim 0.1$ and $\Delta m^2_{12}\lesssim 10^{-3}$~eV$^2$ 
(reactor neutrinos have energies of several MeV, while the most recent experiments
probed $L$ of order 1~km). Second, the solar neutrino puzzle
requires values of $P_{ee}$ significantly different from 1, such that values of  
$\sin^2\theta_{13}$ significantly different from 1 are required 
(see eq.~(\ref{Psol_3})). This, combined with the reactor bound, implies
$\sin^2\theta_{13}\lesssim 0.1$. 

\begin{figure}
\centerline{
\parbox{0.65\textwidth}{\includegraphics[width=0.67\textwidth]{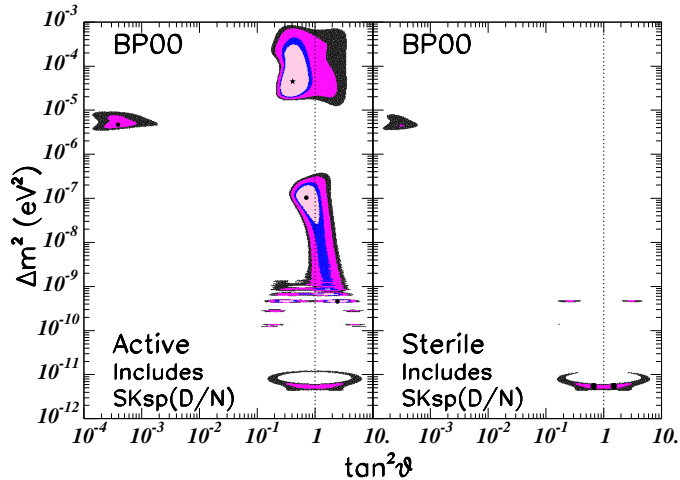}}
\parbox{0.35\textwidth}{\caption{Neutrino oscillation fit to the solar neutrino data,
assuming $\nu_{\mu}\leftrightarrow\nu_{\rm active}$ (left) or 
$\nu_{\mu}\leftrightarrow\nu_{\rm sterile}$ (right). The countors correspond to
90\%, 95\%, 99\%, and 99.7\% (3 sigma) confidence levels. See \cite{BGP} for details.}}
\label{solar_fit}}
\end{figure}

Because the atmospheric data requires 
$|\Delta m^2_{23}|\sim {\rm few}\times 10^{-3}$~eV$^2$,
$\Delta m^2_{23}$--effects for solar neutrinos ``average out'', and 
the solar neutrino oscillations are sensitive to mostly $\Delta m^2_{12}$ and 
$\theta_{12}$ (hence these are referred to as the solar parameters). Currently,
disjoined regions of the parameter space satisfy the data quite well 
\cite{Carlos}, and they are referred
to as: the large mixing angle region, which contains 
$0.2\lesssim\tan^2\theta_{12}\lesssim 3$, $10^{-5}~{\rm eV^2}\lesssim 
\Delta m^2_{12}\lesssim 10^{-3}~{\rm eV^2}$, the low-mass-squared--just-so region,
which contains 
$0.1\lesssim\tan^2\theta_{12}\lesssim 10$, $10^{-10}~{\rm eV^2}\lesssim 
\Delta m^2_{12}\lesssim 10^{-6}~{\rm eV^2}$, and the small mixing angle region,
currently disfavoured with respect to the other two after the publication of the SNO 
\cite{SNO,new_fits,BGP} data, which contains 
$10^{-4}\lesssim\tan^2\theta_{12}\lesssim 10^{-3}$, $10^{-6}~{\rm eV^2}\lesssim 
\Delta m^2_{12}\lesssim 10^{-5}~{\rm eV^2}$. These regions are depicted in 
fig.~\ref{solar_fit} \cite{BGP}. 

Finally, the atmospheric data requires $P_{\mu\mu}$ close to 0.5
for ``large enough'' values of $L/E_{\nu}$, while $P_{ee}$ is roughly 1. The latter is 
easily satisfied given the reactor constraints on $\Delta m^2_{12}$ and 
$\sin^2\theta_{13}$,
while the former determines the values of $|\Delta m^2_{23}|$ and $\theta_{23}$.
Explicitly (ignoring matter effects), 
\begin{equation}
P_{\mu\mu}=1-(\sin^22\theta_{23}\cos^4\theta_{13}+\sin^2\theta_{23}
\sin^22\theta_{13})\sin^2\left(\frac{\Delta_{23}L}{2}\right)+ O(\Delta_{12}L).
\end{equation}
Note that ``maximal mixing'' corresponds to 
$\sin^2\theta_{23}\simeq 0.5/(1-\sin^2\theta_{13})$. Detailed analyses of the atmospheric
data in terms of neutrino oscillations constrain $0.3\lesssim\tan^2\theta_{23}\lesssim3$ and 
$1\times10^{-3}~{\rm eV^2}\lesssim|\Delta m^2_{23}|\lesssim6\times 10^{-3}~{\rm eV^2}$
\cite{Lisi}. 

\section{Including LSND -- Four Neutrino Schemes}
\label{sec:LSND}

The LSND anomaly may be interpreted as evidence for 
$\bar{\nu}_{\mu}\rightarrow\bar{\nu}_{e}$ conversion. If the conversion mechanism is
neutrino oscillations, the LSND result requires $P_{\bar{\mu}\bar{e}}\sim 10^{-3}$ for
neutrino energies of several tens of MeV and a baseline of roughly 
30 metres \cite{LSND}. In order to obtain a small enough oscillation length, 
values of $\Delta m^2\gtrsim 1$~eV$^2$, much larger than the values required
in order to address the solar and atmospheric puzzles, are required. If all three
neutrino puzzles are interpreted in terms of neutrino oscillations, three different
$\Delta m^2$ values are required, which implies the existence of (at least) four
neutrino mass eigenstates. One the other hand, very precise LEP measurements at the 
$Z^0$-boson pole indicate that there are only three neutrino species that couple to the 
$Z^0$-boson. One is forced to conclude, therefore, that the LSND anomaly, combined with 
the solar and atmospheric puzzles, hints at the existence of a fourth neutrino, 
which is a SM singlet (hence referred to as ``sterile,'' $\nu_s$). 
  
There are two rather distinct ways of organising the four neutrino masses-squared
in order to try to solve all the neutrino anomalies. One is the ``3+1'' scheme, which
has four hierarchical masses-squared, $m_1^2<m_2^2<m_3^2<m_4^2$, such that 
$\Delta m^2_{12}$ is responsible for the solar anomaly, 
$\Delta m^2_{23}\simeq\Delta m^2_{13}$ for the atmospheric anomaly, and 
$\Delta m^2_{34}\simeq\Delta m^2_{24}\simeq\Delta m^2_{14}$ for the LSND anomaly
(there are other variations, such as an ``inverted hierarchy'' for the masses 
of $\nu_{1,2,3}$). In
this scheme, the $\nu_4$ state turns out to be predominantly $\nu_s$, and the solar
and atmospheric puzzles are interpreted in terms of (predominantly) active neutrino
oscillations. The LSND anomaly, on the other hand, requires a small $\nu_{e,\mu}$ 
component in $\nu_4$. A reasonable combined fit to all three puzzles exists, and is quite
robust. The biggest obstacle to the 3+1 schemes is to satisfy bounds from terrestrial 
$\nu_{\mu}$ and $\nu_e$ oscillation experiments, which are sensitive to the 
large value of $\Delta m^2$ required to solve the LSND anomaly. For more
details, see the talk by Orlando Peres \cite{Orlando}.  
 
Another possibility is to ``pair up'' the neutrino masses, such that 
$m_3^2\sim m_4^2\gg m_{1,2}^2$. In this case, 
$\Delta m^2_{12}$ is responsible for the solar anomaly, 
$\Delta m^2_{34}$ for the atmospheric anomaly, and 
$\Delta m^2_{13}\simeq\Delta m^2_{14}\simeq\Delta m^2_{23}\simeq\Delta m^2_{24}$ 
for the LSND anomaly (one variation is  
$\Delta m^2_{12}$ responsible for the atmospheric anomaly and 
$\Delta m^2_{34}$ for the solar one). These are referred to as ``2+2'' schemes.
In this case, there are virtually no constraints from terrestrial neutrino oscillation
searches but, instead, one has to worry about properly solving the solar and
atmospheric neutrino puzzles.

The problem is easy to understand. It has to do with the fact that, in a 2+2 scheme,
either solar $\nu_e$ or atmospheric $\nu_{\mu}$ oscillations have to be into a
predominantly sterile state. Explicitely, let us assume $U_{e3}=U_{e4}=0$, and
$U_{\mu1}=U_{\mu2}=0$ (note that these conditions are weakly violated if one
is to solve the LSND anomaly), such that
\begin{eqnarray}
\nu_{1}&=&\cos\vartheta \nu_e+\sin\vartheta (\cos\zeta\nu_{\tau}+\sin\zeta\nu_s), \\
\nu_{2}&=&-\sin\vartheta \nu_e+\cos\vartheta (\cos\zeta\nu_{\tau}+\sin\zeta\nu_s), \\  
\nu_{3}&=&\cos\phi \nu_{\mu}+\sin\phi (-\sin\zeta\nu_{\tau}+\cos\zeta\nu_s), \\
\nu_{4}&=&-\sin\phi \nu_{\mu}+\cos\phi (-\sin\zeta\nu_{\tau}+\cos\zeta\nu_s), 
\end{eqnarray}
In this case, $\vartheta$ ($\phi$) would play the part of the solar (atmospheric) angle,
while $\zeta$ controls the ``amount'' of sterile neutrino in both the solar and the
atmospheric sectors. It is easy to see that if $\zeta=0$, $\nu_{\mu}$ ($\nu_e$) 
oscillates into a pure $\nu_s$ ($\nu_{\tau}$) state, while the situation is reversed
if $\zeta=\pi/2$. 

The current atmospheric data strongly disfavours (at more than 99\% confidence level 
\cite{Atm_talk}) $\zeta=0$, while the solar data disfavours $\zeta=\pi/2$ 
\cite{Solar_talk} (this can also be seen in fig.~\ref{solar_fit}). Detailed quantitative
analysis \cite{Carlos,Lisi,2+2} seem to indicate that, if all atmospheric and solar
data are combined, acceptable fits can be found (the current
best fit point is close to $\sin^2\zeta=0.2$ \cite{2+2}). It is curious to note that,
according to \cite{2+2}, the point where the sterile component is evenly shared
among the solar and atmospheric pairs ($\zeta=\pi/4$) is strongly disfavoured by
the data. As more neutrino data accumulates, it is not impossible to imagine that 
2+2 schemes will be severly cosntrained by the data (perhaps as much, or  
more, than the 3+1 schemes).       

\section{Conclusions and Outlook}
\label{sec:outlook}

It is easy to summarise the results of sec.~\ref{sec:at_work}: 
$|\Delta m^2_{23}|$ and $\theta_{23}$ have been rather well measured, while  
$\Delta m^2_{12}$ and $\theta_{12}$ have also been measured, but rather poorly. Nothing
is known about the value of $\theta_{13}$ (except that it is relatively small), the
value of the CP-odd phase $\delta$, or the neutrino mass-hierarchy (the sign of 
$\Delta m^2_{23}$ in the traditional parametrisation).  

In the near future, the situation is bound to improve significantly, especially in the
solar sector. The on-going solar experiments (in particular SNO) will not only 
establish once and for all that neutrino conversion is the mechanism behind the
solar neutrino problem, but will also improve our knowledge of the solar parameters. 
The near future experiments KamLAND \cite{KamLAND} and 
Borexino \cite{Borexino} 
have the ability to go even further, by not only 
breaking the current degeneracy in the $(\Delta m^2_{12}\times\theta_{12})$-plane but also
determining the solar parameters with unprecedented precision (KamLAND in the case
of the large mixing angle solution \cite{KamLAND,Kam_analysis,Kam_analysis2}, 
Borexino in the case of the low-mass-squared--just-so solution 
\cite{day_night,seasonal}).  
Furthermore, it should also be noted that the KamLAND reactor experiment offers the 
first realistic opportunity to observe a ``real'' oscillatory pattern, while the Borexino
solar neutrino experiment may establish, similar to what SNO will eventually accomplish,
whether there are neutrinos other than $\nu_e$ coming from the Sun \cite{Borex_spectrum}. 

In the ``atmospheric sector,'' the ongoing K2K experiment \cite{K2K}, 
the future MINOS experiment \cite{MINOS},
and the CERN CNGS project \cite{CNGS}
will confirm atmospheric neutrino oscillations, discover
$\tau$-appearance, and extend slightly the sensitivity to $\theta_{13}$, on top of
precisely measuring the atmospheric parameters $|\Delta m^2_{23}|$ and $\theta_{23}$.  

Finally, after the MiniBoone data \cite{MiniBoone}
is collected and analysed, the issue of whether 
the LSND anomaly is indeed a consequence of new neutrino physics will be settled. We 
will then be able to decide whether there is indeed a light sterile neutrino, such
that all three neutrino anomalies are explained by neutrino oscillations. It should
be noted, as briefly discussed in the sec.~\ref{sec:LSND}, that the current neutrino
data already places a significant amount of ``pressure'' on sterile neutrino oscillations,
and the situation is bound to change as new experimental data from solar, reactor
and long-baseline experiments becomes available (for an example, see 
\cite{Kam_analysis2}).  

In five to ten years times, the neutrino mixing matrix will still be far from well known: 
most likely, one of the mixing angles will simply not have been observed (unless
the long-baseline experiments ``get lucky''), and whether or not there is a 
nontrivial
CP-odd phase will remain unknown. Furthermore, we will be unable to say 
whether the neutrinos masses are hierarchical ($m_1^2<m^2_2<m^2_3$), or whether
the hierarchy is inverted ($m_3^2<m_1^2\simeq m^2_2$, such that
$m^2_2-m^2_1\ll m^2_2$). 

These challenges have to be tackled by a new generation of neutrino 
experiments, which requires
cleaner and much more intense neutrino beams, of different flavours (if possible, as
is the case of a neutrino factory). 
 
I conclude by mentioning that there are other ways of probing the neutrino mixing matrix
and mass spectrum. In particular, the future observation of neutrinos from a 
nearby supernova will probably shed light on some of the issues raised above 
\cite{supernova}, while continuing 
searches for neutrinoless double beta decay (which are the best probes of whether the
neutrinos are Majorana or Dirac fermions) and a ``kinematic'' neutrino mass in the 
tritium beta-decay spectrum also add valuable information \cite{Petcov_talk}.  

\section*{Acknowledgements}

It is a pleasure to thank Yoshi Kuno for the invitation and all the organisers of
NuFACT'01 for putting together a well organised and stimulating conference. I also
thank Manfred Lindner for useful conversations.

\end{document}